\documentclass[12pt]{article}
\usepackage{type1cm, amsmath,hyperref}
\usepackage[psamsfonts]{amssymb}
\usepackage[dvips]{graphicx}

\renewcommand{\baselinestretch}{1.3}
\numberwithin{equation}{section}

\newcommand{\pa}{\partial}

\DeclareMathOperator*{\Tr}{{\rm Tr}}

\newcommand{\ti}{\tilde}

\renewcommand{\baselinestretch}{1.2}

\begin{document}


\renewcommand{\baselinestretch}{1.2}

\begin{titlepage}
\renewcommand{\thefootnote}{\fnsymbol{footnote}}
\begin{flushright}
 \begin{tabular}{l}
 UT-07-27 \\
 \end{tabular}
\end{flushright}

 \vfill
 \begin{center}
 \font\titlerm=cmr10 scaled\magstep4
 \font\titlei=cmmi10 scaled\magstep4
 \font\titleis=cmmi7 scaled\magstep4
 \centerline{}
 \vskip .2 truecm
 \centerline{\titlerm Instanton Counting  }
  \vskip .2 truecm
   \centerline{\titlerm and }
  \vskip .2 truecm
 \centerline{\titlerm
  Matrix Model}

 \vskip 1.6 truecm

\noindent{{\large
  Ta-Sheng Tai\footnote{E-mail:tasheng@hep-th.phys.s.u-tokyo.ac.jp} }}
\bigskip
 \vskip .9 truecm
\centerline{\it $^a$ Department of Physics, Faculty of Science,
University of Tokyo}
\centerline{\it Hongo, Bunkyo-ku, Tokyo 113-0033, JAPAN}
\vskip .4 truecm
\end{center}
 \vfill
\vskip 0.5 truecm

\begin{abstract}
We construct an Imbimbo-Mukhi type matrix model, which reproduces
exactly the partition function of ${\mathbb{CP}^1}$ topological
strings in the small phase space, Nekrasov's instanton counting in
${\cal{N}}=2$ gauge theory and the large $N$ limit of the
partition function in 2-dimensional Yang-Mills theory on a sphere.
In addition, we propose a dual Stieltjes-Wigert type matrix model,
which emerges when all-genus topological string amplitudes on certain simple
toric Calabi-Yau manifolds are compared with the Imbimbo-Mukhi
type model.

\end{abstract}

\vfill
\vskip 0.5 truecm

\setcounter{footnote}{0}
\renewcommand{\thefootnote}{\arabic{footnote}}
\end{titlepage}

\newpage

\section{Introduction}

Recently, classical geometries prove to be
very powerful in determining strong coupling gauge theory dynamics.
A familiar story is the AdS/CFT correspondence in ${\cal{N}}=4$ SYM\cite{M}.
Similarly,
computing the confining phase superpotential in ${\cal{N}}=1$
gauge theory has been
facilitated by a large $N$ duality for conifolds, called the geometric transition%
\footnote{See \cite{G} for a recent review.}. On the dual deformed
conifold, period integrals over holomorphic cycles are responsible
for the effective superpotential of gluino
condensates\cite{1,2,3}. In this ${\cal{N}}=1$ context, Dijkgraaf
and Vafa conjectured that the all-genus free energy of certain
target CY's, locally diffeomorphic to $T^\ast S^3$, reduces to a
hermitian (DV) matrix model\cite{4,5,7}. As a further development
of this celebrated matrix/geometry correspondence, for more
general non-compact CY's, Aganagic et al. provided an unifying way
in \cite{aga0}, where the ${\cal{W}}$-algebra symmetry
(holomorphic diffeomorphism of CY's) in disguise is fully
exploited to solve the all-genus free energy. In terms of the
matrix model, these are known as ${\cal{W}}$-constraints on the
$\tau$-function of integrable hierarchies. Adopting the viewpoint
that the partition function on CY 3-folds can be regarded as a
multi-point function of non-compact B-branes, we can find its
connection to the matrix model $\tau$-function via the
Frobenius-Kontsevich-Miwa transformation.

Our situation here is that the full non-perturbative partition
function of ${\cal{N}}=2$
Seiberg-Witten theory%
\footnote{It can be either abelian or non-abelian. The former case
is realized by wrapping a D5-brane on a $\mathbb{CP}^1 \subset
\mathbf{K}3$, where the non-perturbative effect arises from
worldsheet instantons. By rearranging the abelian partition
function, one as well obtains a non-abelian version, see Appendix
B. } is found to be captured by an Imbimbo-Mukhi%
\footnote{The Imbimbo-Mukhi matrix model\cite{IM} is a $\tau$-function
satisfying the 2-time Toda lattice hierarchy.} type matrix model in the `t Hooft
limit. Alternatively, because this instanton partition
function\cite{N1,N2,N3} is precisely that of $\mathbb{CP}^1$
topological strings in the small phase space (see, e.g.
\cite{C1,C2}), our proposal may serve as another sort of
matrix/geometry correspondence, though now the integrable
structure is quite simple.

The outline of this article is as follows. In the next section, we
review the derivation of the Gromov-Witten prepotential of
$\mathbb{CP}^1$ topological strings in the small phase space. We
focus on its free fermion representation. Though the integrable
structure is simple, it is still fruitful enough to exhibit the
underlying Toda hierarchy. In sec.3, we construct an Imbimbo-Mukhi
type matrix model to compute the $\mathbb{CP}^1$ partition
function and sketch a proof using essentially properties of Schur
polynomials. Meanwhile, our result gives spontaneously Nekrasov's
instanton counting in ${\cal{N}}=2$ gauge theory. Comments on
relations to 2-dimensional Yang-Mills theory are summarized in
sec.4. Based on observations of all-genus amplitudes of
topological strings on some simple toric CY's, we also propose a
Stieltjes-Wigert type matrix model dual to the previous
Imbimbo-Mukhi type one in the 't Hooft limit.

\section{$\tau$-function of $\mathbb{CP}^1$ model}

The complex projective space $\mathbb{CP}^1$ is the simplest
target manifold with a positive first Chern class. Since
$\mathbb{CP}^1$ has only two de Rham classes, i.e. the identity
and the K{\"{a}}hler class, the topological A-model on it will
have two corresponding physical observables, i.e. $P$ and $Q$,
respectively. Note that $t_{n,P}$ and $t_{n,Q}$ ($n>0$) indicating
couplings to gravitational descendants of $P$ and $Q$ are
identified with times of the Toda hierarchy in \cite{C2,C3}. When
all couplings are turned off (in the small phase space), the
(genus zero) free energy is
\begin{align}
\frac{1}{2}t^2_{0,P} t_{0,Q} +e^{t_{0,Q}},
\end{align}
where $t_{0,Q}:=t$ denotes
the K{\"{a}}hler parameter of $\mathbb{CP}^1$ and $t_{0,P}$
will
play the role of
the complex scalar vev in the ${\cal{N}}$=2 $U(1)$ vector multiplet.
Besides,
the exponential term comes from the degree-1 instanton.

Following \cite{C1,C4}, the partition function ($\tau$-function)
can be expressed as a sum 
over Young tableaux 
${\lambda}=(k_1,\cdots,k_j,\cdots)$ ($k_i$ : $i$-th row length, $|\lambda|: 
\text{number of boxes}$):
\begin{align}
\begin{aligned}
Z_{}^{CP^1}&=\sum_{\lambda }\frac{\mathbf{m}^2_{\lambda }}
{\hbar^{2|\lambda |}}\exp\big({t\frac{\text{ch}_2(a,\lambda )}{2\hbar^2}}\big),
~~~~~~~a=t_{0,P},\\
\mathbf{m}_{\lambda }&=\prod_{\square}\frac{1}{h(\square)}=
\frac{d_{\lambda }}{|\lambda |!}=\prod_{i<j}
\frac{k_i-k_j+j-i}{j-i}, ~~~~~~~h(\square): \text {hook length}.
\label{cp}
\end{aligned}
\end{align}
Here, $\hbar$ denotes the genus expansion parameter. Note that
$\mathbf{m}_{\lambda }$ and $d_{\lambda }$, respectively, are the
Plancherel measure and the dimension of $\lambda $ as a
representation of $S_{\lambda }$ (symmetric group). In addition,
the Chern polynomial $\text{Ch}_n$ is defined as a Fourier transform of the
second derivative of $f_{\lambda }(x,a)$, which is the $profile$
$function $ obtained by rotating counterclockwise
 the Young tableau
by $\frac{4}{3}\pi$:
\begin{align}
\frac{1}{2}\int dx ~f''_{\lambda }(x,a)~e^{ux}=\sum^{\infty}_{n=0}
 \frac{u^n}{n!}\text{ch}_n (a,\lambda ),
\end{align}
where (box size: $\hbar\times\hbar$)
\begin{align}
\begin{aligned}
f_{\lambda}(x)&=|x-a|+\rho (x),\\
\rho(x)&=\sum^{\infty}_{i=1} \Big(
|x-a-\hbar k_i +\hbar(i-1)|-
|x-a-\hbar k_i+\hbar i |\\
&~~~~~~~~~+|x-a+\hbar i|-|x-a+\hbar(i-1)|
\Big).
\label{fff}
\end{aligned}
\end{align}

Alternatively, $Z_{}^{CP^1}$ can be written into a more compact
matrix element
form:
\begin{align}
Z_{}^{CP^1}=\langle0  | e^{\frac{J_{1}}{\hbar}}~ e^{{tL_0}}~e^{\frac{J_{-1}}{\hbar}}
|0 \rangle .
\end{align}
To explain the notation, let us make use of fermionic operators $b,c$ satisfying
the usual properties:
\begin{align}
\{b_r , c_s\}=\delta_{r+s,0}, ~~~~~r,s\in\mathbb{Z}+1/2;~~~~~~~~~~~
b_r|0\rangle =c_s|0\rangle =0, ~~~~~\forall r,s>0
\label{1}
\end{align}
to define an orthogonal basis \{$|\lambda \rangle$\} w.r.t.
$\lambda$:
\begin{align}
|\lambda \rangle =\prod^\infty_{i=1} b_{i-k_i -\frac{1}{2}}|0'\rangle, ~~~~~
\langle \lambda |\kappa \rangle =\delta_{\lambda ,\kappa };~~~~~~~~~~~
c_r|0'\rangle =0,~~~~~\forall r ,
\end{align}
which satisfies
\begin{align}
e^{\frac{J_{-1}}{\hbar}}|0\rangle = \sum_{\lambda }
\frac{1}{\hbar^{|\lambda |}\prod_{\square\in\lambda } h(\square)
}|\lambda \rangle
\label{use}
\end{align}
and
\begin{align}
L_0 |\lambda \rangle =\frac{1}{2\hbar^2 }\text{ch}_2 (a,\lambda )
|\lambda \rangle .
\end{align}
Note that the Virasoro generator $L_0=\frac{1}{2}J_0^2 + J_{-1} J_1$
consists of the $U(1)$ current defined using the bosonization:
\begin{align}
J(z)=:b(z)c(z):=\sum_{n\in\mathbb{Z}} J_n z^{-n-1}, ~~~~
J_n=\sum_{s\in \mathbb{Z}+1/2}:b_s c_{n-s}:,~~~~[J_m , J_{-n}]=n\delta_{m+n,0}.
\label{2}
\end{align}
For $J_0=\frac{a}{\hbar}$, the matrix element reads
\begin{align}
Z_{}^{CP^1}=\exp \frac{1}{\hbar^2}\Big(e^{t} + \frac{1}{2}t{a^2}\Big).
\end{align}
It can be verified that $(\hbar=1)$
\begin{align}
\frac{\pa^2}{\pa t^2}\log Z^{CP^1}(a)=
\frac{ Z^{CP^1}(a-1)~ Z^{CP^1}(a+1)}{ \big( Z^{CP^1}(a) \big)^2},
\end{align}
which is none other than
the first equation of the Toda lattice hierarchy.

\section{Imbimbo-Mukhi type matrix model}

Before writing down our matrix model, let us first study an
illuminating generalization, namely, the Imbimbo-Mukhi matrix
model which describes the deformed conifold $T^\ast S^3$ with
B-brane insertions.

According to \cite{aga0}, on the B-model side, to vary the complex
structure of CY's is done by wrapping non-compact B-branes on
degenerating fibers. The problem hence boils down to dealing with
the moduli space of a Riemann surface. Moreover, with the special
geometry relation, quantizing the complex moduli is realized by
introducing quantum Kodaira-Spencer fields. The vacuum expectation
value of the KS field in turn defines a $quantum$ Riemann surface
which determines the CY with varied complex moduli. For full
details on this subject, readers are recommended to consult
\cite{aga0}.

We quickly review the case of $T^\ast S^3$, whose defining
equation is
\begin{align}
uv-H(x,y)=0, ~~~~~~~~~~H(x,y) =xy-\mu.
\label{sim}
\end{align}
Two KS fields (chiral bosons)
\begin{align}
\begin{aligned}
x=\partial_y \ti\phi (y), ~~~~~~~~~~
y=-\partial_x \phi (x)
\end{aligned}
\end{align}
are defined on asymptotic regions $x,y\to\infty$ of the degenerating locus
$H(x,y)=0$ which has a topology of the sphere.
If there are initially $N$ non-compact
B-branes at $x$-patch
\begin{align}
\langle N| \psi (x_1) \cdots \psi (x_N) |W\rangle, ~~~~~~~~
\psi=e^{\frac{1}{g_s}\int y},
\label{WW}
\end{align}where
$| W\rangle \in{\cal{H}}$ (${\cal{H}}$: Hilbert space of chiral
bosons $\phi$) and $\langle N|$ denotes an $N$-fermion state, the
complex structure of $T^\ast S^3$ is deformed to be
\begin{align}
xy=\mu ~~\longrightarrow~~ x(y+\sum_{n>0} nt_n x^{n-1})=\mu,
~~~~~~~~t_n=-\frac{g_s}{n}\sum^N_{i=1} x_i^{-n}.
\label{dg}
\end{align}
Notice that the canonical transformation relating $x$-patch and
$y$-patch gives rise to the $S$-matrix, which acts on the B-brane
operator as a Fourier transform, i.e.
\begin{align}
\ti\psi(y)=S\psi(y)=\frac{1}{\sqrt{2\pi g_s}}\int dx ~e^{-\frac{xy}{g_s}} \psi(x).
\label{S}
\end{align}
In the deformed geometry \eqref{dg}, moving B-branes
from $x$-patch
to $y$-patch $Y=\text{diag}(y_1,\cdots,y_N)$ leads to
\begin{align}
\int \prod_{i=1}^N dx_i ~e^{-\sum^N_{i=1} x_i y_i/g_s} \langle
\psi (x_1) \cdots \psi (x_N) \rangle. \label{KKK}
\end{align}
Further, via the Harish-Chandra-Itzykson-Zuber formula
\begin{align}
\int DU ~ e^{i\Tr (BX )} \propto\frac{\det
e^{ i x_j p_j}}{\triangle(x)\triangle(p)},   ~~~~~~~~~~\triangle: \text{VanderMonde},
\label{IZ}
\end{align}
where $X=\text{diag}(x_1,\cdots,x_N)$ and $B=U^\dagger P U$ with
${{P}}=\text{diag}(p_1,\cdots,p_N)$, we are able to obtain a
Kontsevich-like matrix model for \eqref{KKK}, i.e.
\begin{align}
\begin{aligned}
Z(\ti t, t)&=\frac{1}{\triangle(y)}
\int \prod dx_i~\triangle(x)
(\det X)^{-\mu/g_s}e^{\frac{1}{g_s}(\sum_n t_n \Tr X^n +  \sum_i x_i y_i)}\\
&=\int_{N \times N} dX
(\det X)^{-\mu/g_s}e^{\frac{1}{g_s}\Tr(\sum_n t_n \Tr X^n +  XY)},
\label{Tau}
\end{aligned}
\end{align}
where $\ti t_n=\frac{g_s}{n}\Tr Y^{-n}$. For large enough $N$, 
we choose $\mu=g_s N$. \eqref{Tau} is the so-called Imbimbo-Mukhi
matrix model first derived in \cite{IM}.

\subsection{Proof}
We are now in a position to propose our model.
Under the 't Hooft limit:
\begin{align}
\mu=Ng_s: ~\text{large but fixed}, ~~~~~~~~~~N\to \infty, ~~~~~~~~~~g_s \to 0,
\end{align}
the following hermitian matrix integral reproduces Nekrasov's instanton
counting formula as well as the previous $Z^{CP^1}$, i.e.
\begin{align}
\mathbf{Z}(\mathbf{U})=\int_{N \times N} dX (\det X)^{-\mu} \exp
\Tr {\mu} \Big( \sum_{n>0} \mathbf{K}_n X^n +  X \mathbf{U}^{-1}
\Big), \label{type}
\end{align}
where
\begin{align}
 \mu\mathbf{K}_n=-\frac{1}{n}\Tr \mathbf{U}^n, ~~~~~~~~~~~
\mathbf{U}=\text{diag}~(q^\frac{1}{2},\cdots,q^{N-\frac{1}{2}}),
~~~~~~~~~~~q=e^{-g_s}.
\end{align}

Let us now sketch a proof of our proposal. A fact of particular
use is that the $\tau$-function \eqref{Tau} admits a $S$-matrix
formulation\cite{DPM}, where the $S$-matrix has been explicitly
diagonalized thanks to tachyon reflection amplitudes in the
bosonic $c=1$ string. Based on those, we rewrite
$\mathbf{Z}(\mathbf{U})$ into ($S$: $S$-matrix)
\begin{align}
\mathbf{Z}(\mathbf{U})=\langle \ell|S|\ell\rangle , ~~~~~~~~~
|\ell\rangle=\exp \sum^\infty_{n=1}\frac{\Tr \mathbf{U}^n}
{n}\alpha_{-n}|0\rangle ,~~~~~~~~~
\ell_n=\Tr \mathbf{U}^n.
\label{111}
\end{align}
Given all mathematical details in Appendix A, we 
insert the identity operator $1=\sum_R |R\rangle \langle R|$ into \eqref{111} to get
\begin{align}
\mathbf{Z}(\mathbf{U})=\sum_{R} {\cal{N}}_R ~s_R (\ell)
s_R (\ell), ~~~~~~~~~~~~~~~~~{\cal{N}}_R =
\prod_{\square(i,j)}(\mu -i+j),
\end{align}
where $s_R (\ell)$ is the Schur polynomial and the diagonal
element ${\cal{N}}_R$ can be found in \cite{aga}%
\footnote{Note that the diagonal element ${\cal{N}}_R$ also 
appears in the context of ${1}/{2}$ BPS correlators 
in ${\cal{N}}=4$ SYM, see \cite{J,T}.}. Upon imposing
the 't Hooft limit, $\mathbf{Z}(\mathbf{U})$ reduces to a MacMahon
function\cite{M1,NA}%
\footnote{The subleading term $\frac{\kappa (R)}{2}
\mu^{|R|-1}$ in ${\cal{N}}_R$, where $2\sum_{\square(i,j)} (-i+j)
=\kappa (R)$, is omitted when compared to the leading one. } :
\begin{align}
\mathbf{Z}(\mathbf{U}) = \sum_{R} \mu^{|R|} ~s_R (\ell)^2
=\prod^{N=\infty}_{n=1}\frac{1}{(1- \mu q^n)^n} .
\label{MAC}
\end{align}
We have taken advantage of the Cauchy's formula
\begin{align}
\sum_{\lambda} \Tr_\lambda \mathbf{A}~ \Tr_\lambda
\mathbf{B}=\prod_{i,j}\frac{1}{1-a_i b_j},
\end{align}
where $a_i$ ($b_i$) denotes the eigenvalue of $\mathbf{A}$ $(\mathbf{B})$.
Another way defining the Schur polynomial in terms of $q$ is
\begin{align}
s_R (\ell)=\Tr_R \mathbf{U}=q^{-\frac{1}{2}\sum_{(i,j)\in\lambda}(j-i)}\prod_{(i,j)\in\lambda}
\frac{1}{q^{-h(i,j)/2}-q^{h(i,j)/2}},~~~~~~~~~~h(i,j): \text{hook length}.
\label{wei}
\end{align}
By expanding $q$ in \eqref{wei} around small $g_s$,
together with \eqref{cp}, it is seen that
\begin{align}
\sum_R \mu^{|R|} ~s_R (\ell) ^2  ~~\to ~~\sum_R
\big(\frac{\mu}{g^2_s}\big)^ {|R|}~{\mathbf{m}}_{R}^2 . \label{Sc}
\end{align}
Amazingly, the partition function of
${\mathbb{CP}^1}$ in the small phase space is reproduced upon
\begin{align}
g_s\to \hbar, ~~~~~~~~~~~~~~~~~\mu\to e^{t}+\frac{1}{2}a^2 t.
\end{align}
It is fine that $a$ can be turned off to zero because it stands
for the complex scalar vev in the ${\cal{N}}$=2 $U(1)$ vector
multiplet below.

\subsection{Nekrasov's formula}
As is shown, our
Imbimbo-Mukhi type matrix model proves to give
\begin{align}
\sum_\lambda \big(\frac{\mu}{\hbar^2}\big)^
{|\lambda|}~{\mathbf{m}}_{\lambda }^2
\label{res}
\end{align}
under the 't Hooft limit.
As a matter of fact, \eqref{res} is
nothing but Nekrasov's partition function in ${\cal{N}}=2$ $U(1)$
gauge theory\cite{C1}%
\footnote{How to convert \eqref{res} into a non-abelian instanton
partition function is summarized in Appendix B. }:
\begin{align}
\begin{aligned}
Z^{inst} (a,\Lambda, \hbar) &=\sum_{\lambda } \exp (-{\cal{E}}[f,\lambda ]),\\
{\cal{E}}[f,\lambda ]&= \frac{1}{2}\int \int_{x_1 > x_2} dx_1 dx_2
~f_{\lambda}''(x_1)f_{\lambda}''
(x_2)\gamma _{\hbar} (x_1 -x_2,\Lambda ),
\label{inst}
\end{aligned}
\end{align}
where
\begin{align}
\begin{aligned}
\gamma _{\hbar} (x,\Lambda ):&=\frac{d}{ds}\Big|_{s=0}
\frac{\Lambda^s}{\Gamma (s)}\int^\infty_0 \frac{dt}{t^{1-s}}
\frac{e^{-tx}}{(e^{\hbar t}-1)(e^{-\hbar t}-1)}, \\
\Gamma (s)&=\int_0^\infty dt ~e^{-t}~ t^{s-1}.
\end{aligned}
\end{align}
In addition, $\gamma _{\hbar} (x,\Lambda )$ has a $genus$ expansion w.r.t. $\hbar$:
\begin{align}
\begin{aligned}
\gamma _{\hbar} (x,\Lambda )&=\sum_{g=0}^{\infty}\hbar^{2g-2}\gamma_g (x,\Lambda ) \\
&=\frac{1}{\hbar^2} \Big(\frac{1}{2}x^2
\log(\frac{x}{\Lambda })-\frac{3}{4}x^2 \Big)-\frac{1}{12}\log\frac{x}{\Lambda }
+\sum_{g=2}^{\infty}\frac{B_{2g}}{2g(2g-2)}(\frac{\hbar}{x})^{2g-2},
\end{aligned}
\end{align}
where an identity
\begin{align}
\begin{aligned}
\frac{1}{(e^t -1)(e^{-t}-1)}=\frac{d}{dt}\frac{1}{e^t -1}=
\frac{1}{t^2}+\sum^\infty_{g=1} \frac{B_{2g}}{2g(2g-2)!}t^{2g-2}
\end{aligned}
\end{align}and
the definition of the Bernoulli number
\begin{align}
\begin{aligned}
\frac{x}{e^x -1}=\sum_{n=0}^\infty \frac{B_n}{n!}x^n~~~~~~~
\end{aligned}
\end{align}
have been used. Through the substitution
\begin{align}
\mu= \frac{1}{2}a^2 \log \Lambda + \Lambda ,~~~~~~~~~~~~~
\end{align}
we yield\cite{C1}
\begin{align}
\begin{aligned}
Z^{inst}(a,\Lambda ,\hbar)=\sum_\lambda \big(\frac{\mu}{\hbar^2}\big)^
{|\lambda|}~{\mathbf{m}}_{\lambda }^2 .
\label{df}
\end{aligned}
\end{align}
Notice again that choosing $a=0$ does not lose any generality.

\section{Comments}
So far, we have constructed an Imbimbo-Mukhi type matrix model. In
the 't Hooft limit, it reproduces the partition function of
$\mathbb{CP}^1$ topological strings in the small phase space.
Besides, it is none other than Nekrasov's instanton counting
formula in this limit.

Let us include two more applications. The first is that our model
can recover the partition function of 2-dimensional Yang-Mills
theory on a sphere. Another aspect is that the all-genus free
energy of special toric Calabi-Yaus can be derived from our model.
To this end, we further propose a dual Stieltjes-Wigert type
matrix model.
\subsection{2-dimensional Yang-Mills}

The partition function of 2d $U(N)$ YM on a genus $g$ and area $A$
orientable manifold is\cite{RR,RR1}
\begin{align}
\begin{aligned}
Z^{2d}=\sum_R (dim R)^{2-2g} \exp \Big( -\frac{\lambda A}{2N}C_2 (R) \Big),
\end{aligned}
\end{align}
where $\lambda$ is the 't Hooft coupling of 2d YM, $C_2 (R)$ is
the quadratic Casimir of the representation $R$ (conventions
inherited from \eqref{cp} and \cite{ohta}; the box size is
$\hbar\times \hbar$)
\begin{align}
C_2 (R)= \sum^N_{i=1} \big( Nk_i +k_i(k_i-2i+1) \big)
=\int dx f''_R (x)~ (\frac{N}{4\hbar^2}x^2 + \frac{1}{6\hbar^3}x^3)
\end{align}
and
\begin{align}
\begin{aligned}
dim R&=\prod_{1\le i<j\le N}
\frac{k_i-k_j+j-i}{j-i}\\
&=\exp \Big( -\frac{1}{4}\int \int _{x>y} ~dx dy
f_R''(x) f_R''(y) ~\gamma_{\hbar}(x-y) +\frac{1}{2}\int dx f''_R (x) ~
\big(\gamma_\hbar (x+\hbar N)-\gamma_\hbar (\hbar N)\big) \Big).
\end{aligned}
\end{align}
Here, we have set $a=0$ and $\Lambda =1$ for $f_R (x,a)$ and
$\gamma_{\hbar} (x,\Lambda)$. Note that the second term in the
above second line serves as a cut-off which will be dropped out
when $N$ is taken to infinity.

At the first sight, when $N$ goes to infinity,
$dimR\to\mathbf{m}_R$ and $C_2 (R)\sim N|R|$, so the genus zero
$Z^{2d}$ is exactly of the form \eqref{res}. Let us check this
point precisely. Combining everything, we yield
\begin{align}
\begin{aligned}
Z^{2d}=
\sum_R
\exp \Big( &-\frac{2-2g}{4}\int \int _{x>y} ~dx dy
f_R''(x) f_R''(y) ~\gamma_{\hbar}(x-y) \\
&+\frac{2-2g}{2}\int dx f''_R (x) ~
\big(\gamma_\hbar (x+\hbar N)-\gamma_\hbar (\hbar N)\big)
-\frac{\lambda A}{4N}
\int dx f''_R (x) ~(\frac{N}{4\hbar^2}x^2 + \frac{1}{6\hbar^3}x^3)\Big).
\end{aligned}
\end{align}
For large $N$,
we turn out to get
\begin{align}
\begin{aligned}
Z^{2d}\to
\sum_R
\exp \Big( &-\frac{1-g}{2}\int \int _{x>y} ~dx dy
f_R''(x) f_R''(y) ~\gamma_{\hbar}(x-y) -\frac{\lambda A}{8\hbar^2}
\int dx f''_R (x) ~x^2 \Big).
\end{aligned}
\end{align}
Utilizing another equality for the profile function
$f_{R}(x)$\cite{N2}:
\begin{align}
\begin{aligned}
|R|=
\frac{1}{4\hbar^2} \int dx f_R''(x) ~x^2,
\end{aligned}
\end{align}
in the genus zero case, we are left with
\begin{align}
\begin{aligned}
Z^{2d}=\sum_R \exp \Big(\frac{-\lambda A|R|}{2}\Big) ~\frac{\mathbf{m}_R^2}{\hbar^{2|R|}}.
\end{aligned}
\end{align}
In other words, we may have recovered the 2d YM partition function in
large $N$ limit using our matrix model upon identifying%
\footnote{Note that both $\mu$ and $e^{-\lambda A/2}$ can be
complex. For $\mu$, this is true in the context of the CY
manifold. In the case of ${-\lambda A/2}$, it plays the role of the gauge
coupling of a 4-dimensional theory (e.g. D5-branes compactified on
the $S^2$ of a resolved conifold) and can be complexified by
turning on the theta angle.} $\mu\equiv e^{-\lambda A/2}$.

\subsection{Toric Calabi-Yau's and unknot Wilson loops}
It is well-known that
$U(N)$ level $\ell$ Chern-Simons gauge theory on $S^3$ is large $N$
dual to
the topological A-model on a resolved conifold ${\cal{O}}(-1)+
{\cal{O}}(-1)\to {\mathbb{P}^1}$.
That is, the partition function on the resolved conifold
(K{\"{a}}hler
parameter $t=g_s N$, $q=e^{-g_s}$)
\begin{align}
\begin{aligned}
Z_{P^1}=\exp (-{\cal{F}}),~~~~~~~~{\cal{F}}=\sum_{n>0} \frac{e^{-nt}}{n \Big(
2\sin(ng_s/2) \Big)^2}, ~~~~~~~~g_s =\frac{2\pi i}{\ell+N}
\end{aligned}
\end{align}
is that of Chern-Simons under the 't Hooft expansion\cite{G1,G2}.
This is an open/closed duality realized via the geometric transition of
conifolds.

As another example\cite{M1}, let us consider two $U(N) $
Chern-Simons on two $S^3$'s, separated from each other by a
complexified K{\"{a}}hler parameter $t'$. The Ooguri-Vafa
operator\cite{OV} associated with two holonomies in this setup is
\begin{align}
\begin{aligned}
Z(U,V,t')=\exp \big( \sum_{n>0} \frac{1}{n} e^{nt'} \Tr U^n \Tr V^n \big)
=\sum_R  e^{-t'|R|} \Tr_R U \Tr_R V
\end{aligned}
\end{align}
with the normalized vev
\begin{align}
\begin{aligned}
\langle Z(U,V,t')\rangle
&=\sum_R  e^{-t'|R|} W_R ({\cal{K}}) W_R ({\cal{K}})=\sum_R
M^{|R|} s_R(x_i = q^{\frac{1}{2}-i})^2,\\
W_R ({\cal{K}})&=s_R (x_i=q^{(N-2i+1)/2}), ~~~~~~~~~M=e^{-t'}q^N.~~~~~~
~~~
\label{555}
\end{aligned}
\end{align}
Note that $W_R ({\cal{K}})$ stands for an unknot Wilson loop in
$S^3$. After the geometric transition, a smooth toric CY emerges
which is characterized by three K{\"{a}}hler parameters, namely,
$t_1=t_2=g_s N$ and $t=t'-(\frac{t_1 +t_2}{2})$. The free energy
$-\log\langle Z(U,V,t')\rangle$ is
\begin{align}
{\cal{F}}=\sum_{n>0} \frac{e^{-nt_1}+e^{-nt_2}+
e^{-nt}(1-e^{-nt_1})(1-e^{-nt_2})}{n \Big(
2\sin(ng_s/2) \Big)^2}.
\label{fff}
\end{align}
It is found that \eqref{555} readily agrees with our result in \eqref{Sc} if 
$q\to q^{-1}$ and $M\to \mu$ are carried out in \eqref{555}.

\subsection {Stieltjes-Wigert type matrix model}
Inspired by \eqref{fff}, we are capable of proposing a Stieltjes-Wigert type matrix model which is dual to \eqref{type}.
That is,
\begin{align}
\begin{aligned}
Z_{SW}=\frac{1}{\text{vol}U(N)}&\int_{N\times N} dY \exp \Tr \Big(
-\frac{1}{2g_s} (\log Y)^2 + \log (Y
\otimes 1_{m \times m}
-1_{N\times N}\otimes
\mathbf{V})   \Big), \\
\mathbf{V}&= \text{diag}(v_1,\cdots,v_i,\cdots), ~~~~~~~~~~~~i=1,\cdots,m,
\label{sss}
\end{aligned}
\end{align}
where $Y$ is hermitian and the measure gives the usual
VanderMonde. Without the second logarithm term inside the potential, \eqref{sss} is
originally designated\cite{M2,M3,F,T1}%
\footnote{ See \cite{T2,T3} 
for more details about Chern-Simons theory and 
Stieltjes-Wigert polynomials. }
to rewrite the partition
function of $U(N)$ Chern-Simons theory on $S^3$.

To prove the equivalence to \eqref{type}, we note that \eqref{sss}
dictates a B-model calculation of the non-compact B-brane
correlator\cite{OK,H} with their moduli $\mathbf{V}$. Put it
differently, \eqref{sss} is a mirror description of the A-model
resolved conifold, on whose external leg of the toric diagram a
stack of probe Lagrangian branes is inserted.
Besides, each component of $\mathbf{V}$ is chosen to be 
$v_i=\exp -g_s (\ell-i+m+\frac{1}{2})$%
\footnote{As pointed out in \cite{O1}, $\ell$ and $m$ here assign
an $U(N)$ representation via a 
rectangular Young tableau of $m$ rows and $\ell$ columns. }, which immediately 
implies that these Lagrangian branes are all located at the
same place and, therefore, trigger a geometric transition such
that a new toric CY appears, referred to as a bubbling CY in
\cite{O1}.

Translating everything into the A-model language, one can utilize
the melting crystal method to write down the Lagrangian brane
scattering amplitude via the quantum dilogarithm $L(v,q)$ 
as (K{\"{a}}hler parameter of $\mathbb{P}^1$: $t=(N+m)g_s$)
\begin{align}
\begin{aligned}
Z_{crystal}&=Z_{P^1}(t) M(q) \prod_{i<j} (1-\frac{v_i}{v_j}) \prod^{m}_{i=1}
\frac{L(v_i, q)}{L(v_i e^{-t},q)}, \\
L(v, q)&=\prod^{\infty}_{k=1} (1-vq^k)=
\exp\Big(\sum^{\infty}_{k=1} \frac{v^{k}}{k[k]} \Big), ~~~~~~~~~~[k]=q^{k/2} -q^{-k/2},
\label{MMM}
\end{aligned}
\end{align}
which is essentially the same as \eqref{sss} up to an irrelevant
overall factor and a framing choice\cite{H}. Moreover, by doing so
and inserting the moduli variable $\mathbf{V}$, it is seen that
$-\log \big(Z_{crystal}/M(q)^2 \big)$ coincides with \eqref{fff},
except now\cite{O1}
\begin{align}
t_1= g_s m, ~~~~~~~t_2=g_s (N-m), ~~~~~~~t=g_s \ell.
\label{qqqq}
\end{align}
In other words, given the result discussed 
in sec.4.2, the Stieltjes-Wigert type matrix model in \eqref{sss} can be
regarded to be equivalent to \eqref{type} (up to MacMahon's) in the 
't Hooft limit through 
$m=\frac{N}{2}=N'\to \infty$, $q\to q^{-1}$ and $e^t \to \mu $.

\subsection*{Acknowledgements}

The author would like to thank Feng-Li Lin, Yutaka Matsuo,
Kazuhiro Sakai, Yuji Sugawara, Ryo Suzuki, Masato Taki, Satoshi
Yamaguchi for helpful discussions and, especially, Tohru Eguchi for
reading carefully the manuscript. Also, he gratefully
acknowledges the hospitality of the string group of National
Taiwan University, where the final stage of this work was done. This work is partly supported by the
21COE program of University of Tokyo.

\appendix
\section{Conventions}
Since we heavily rely on the $U(\infty)$ representation theory and
symmetric polynomials, let us explain briefly the notation
following \cite{M1}.
Recall that, via the Frobenius formula, the
Schur polynomial labeled by a Young tableau $R$ is expressed as (for an $N\times N$ unitary matrix $U$)
\begin{align}
 \langle R|U\rangle =\Tr_R U =\sum_{\vec{k}}
\frac{\chi_R (C(\vec{k}))}{z_{\vec{k}}}\Upsilon_{\vec{k}}(U)=
\frac{\det u^{k_i +N-i}_j}{\det u_j^{N-i}},
\label{412}
\end{align}
where $u_i$'s are eigenvalues of $U$, while $k_i$ is the $i$-th row length
of $R$. In addition, $\chi_R (C(\vec{k}))$ is the character w.r.t. $R$
evaluated at the conjugacy class $C(\vec{k})$
of the symmetric group $S_l$
where
\begin{align}
l=\sum_{j}j~k_j,~~~ ~~~~~~~\vec{k}=(k_1,k_2,\cdots)
\end{align}
and
\begin{align}
\Upsilon_{\vec{k}}(U)=\prod_{j=1}^\infty (\Tr U^j)^{k_j}, ~~~ ~~~~~~~
z_{\vec{k}}=\prod_j k_j !~j^{k_j}.
\end{align}
The character $\chi_R (C(\vec{k}))$ is referred to as $\langle
R|\vec{k} \rangle$ if one quotes the 2d CFT techniques in
\eqref{1} and \eqref{2} to define what follow:
\begin{align}
\begin{aligned}
|\vec{k}\rangle &= \prod^\infty_{j=1}(\alpha_{-j})^{k_j} |0\rangle ,
~~~~~~~|R\rangle =
\epsilon (R)\prod^r_{i=1} b_{-m_i-\frac{1}{2}}c_{-n_i -\frac{1}{2}}|0\rangle
,\\
\sum_{\vec{k}}\frac{1}{z_{\vec{k}}} |{\vec{k}}\rangle \langle {\vec{k}}| &= 1,
~~~~~~~\langle R|R'\rangle =\delta_{R,R'},~~~~~~~
\epsilon (R)=\exp{i\pi}\Big( \sum^r_{i=1} n_i + \frac{1}{2}r(r-1) \Big),
\end{aligned}
\end{align}
where $r$ is the number of diagonal boxes of $R$,
while integers $m_i$ and $n_i$ $(i=1,\cdots,r)$ are
\begin{align}
m_i=h_i - i,~~~~~~~~~~~~~~~~~n_i=v_i - i.
\end{align}
Here, $h_i$ ($v_i$) labels the $i$-th row (column) length. These
tools altogether facilitate another expression for a Schur polynomial, i.e.
\begin{align}
\begin{aligned}
s_R (\omega)=\Tr_R \mathbf{W} =\langle R|\omega \rangle ,~~~~~~~~~|\omega \rangle =
\exp \sum^\infty_{n=1}\frac{\omega_n}{n}\alpha_{-n}|0\rangle
,~~~~~~~~~\omega_n=\Tr \mathbf{W}^n.
\label{suh}
\end{aligned}
\end{align}

\section{Non-abelian instanton counting}
For ${\cal{N}}=2$ $SU(m)$ gauge theory, the non-perturbative part
of Nekrasov's formula is
\begin{align}
Z^{int}=\sum_{\vec{R}} \Lambda^{|\vec{R}|}
\prod_{(s,i)\ne(r,j)}\frac{a_s-a_r+
\hbar(k_{s,i}-k_{r,j}+j-i)}{a_s -a_r +\hbar (j-i)},
~~~~~~~~~~{|\vec{R}|}=\sum^{m}_{i=1} |\vec{R}_i|,
 \label{c1}
\end{align}
where $\Lambda $ is the dynamical scale and $\vec{R}$ consists of
$m$ Young tableaux located at $a_r$'s\cite{N1}. According to \cite{ohta},
given an arbitrary Young tableau $\lambda $, we can divide it into
$m$ subpartitions $\subset \vec{R}$ and convert its Plancherel
measure into \eqref{c1} as follows:
\begin{align}
\begin{aligned}
\frac{1}{\hbar^{2|\lambda|}} \mathbf{m}^2_{\lambda }
&=\lim_{N\to\infty}\prod^{N}_{i<j}\Big( \frac{k_i-k_j+j-i}{j-i}\Big)^2\\
&=\prod^m_{s=1}\prod^{N_s}_{i\ne j}\frac{k_{s,i}-k_{s,j}+j-i}{j-i}
\prod^m_{s\ne r}\prod^{N_s}_{i=1}\prod^{N_r}_{j=1}
\frac{W_s-W_r+k_{s,i}-k_{r,j}+j-i}{W_s -W_r +j-i}\\
&=\prod_{(s,i)\ne(r,j)}\frac{a_s-a_r+
\hbar(k_{s,i}-k_{r,j}+j-i)}{a_s -a_r +\hbar (j-i)},\\
\hbar W_s&=\hbar \big(N-(N_1 +\cdots+N_{s-1})\big)={a_s},
\end{aligned}
\end{align}
where $a_r$'s denote vevs of complex scalars in vector multiplets of
unbroken $U(1)^{m-1} \subset SU(m)$.

\baselineskip=14pt

\end{document}